\documentclass{article}

\usepackage{arxiv}
\usepackage[utf8]{inputenc} 
\usepackage[T1]{fontenc}    
\usepackage{hyperref}       
\usepackage{url}            
\usepackage{booktabs}       
\usepackage{amsfonts}       
\usepackage{nicefrac}       
\usepackage{microtype}      
\usepackage{graphicx}
\usepackage[numbers]{natbib}
\usepackage{doi}
\usepackage{multirow} 
\usepackage{amssymb}
\usepackage{pifont}

\title{Early-Stage Requirements Transformation Approaches: A Systematic Review}

\author{ \href{https://orcid.org/0000-0003-4355-7987}{\includegraphics[scale=0.06]{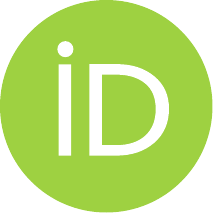}\hspace{1mm}Keletso J.~Letsholo} \\
	Faculty of Computer Information Science \\
	Higher Colleges of Technology\\
	Abu Dhabi, UAE\\
	\texttt{kletsholo@hct.ac.ae} \\
}

\date{}

\renewcommand{\headeright}{}
\renewcommand{\undertitle}{}

\hypersetup{
pdftitle={A template for the arxiv style},
pdfsubject={cs.SE, cs.AI},
pdfauthor={Keletso J.~Letsholo},
pdfkeywords={Textual Requirements, Analysis Models, Model Transformation},
}

\begin{document}
\maketitle

\begin{abstract}

Transformation approaches for automatically constructing analysis models from textual requirements are critical to software development, as they can bring forward the use of precise formal languages from the coding phase to the requirement analysis phase in the software development life-cycle. Over the decades, numerous transformation approaches have been developed in an attempt to fully or partially automate this initial phase.
This systematic review examines transformation approaches in the early stages of software development, examining 25 studies on early-stage requirements transformation documented between 2000 and 2014. The review highlights the widespread use of natural language processing techniques, with tools like the Stanford parser and WordNet being essential. Intermediate models are often used in the transformation process to bridge the gap between textual requirements and analysis models.
Significant advancements have been made in early-stage requirements transformation approaches; however, several areas require attention to enhance their effectiveness and reliability. A challenge identified is the lack of robust evaluation methods, with most approaches using simple case studies and running examples for evaluation. This makes it difficult to compare and evaluate the performance these approaches.
Although most approaches can generate structural models from textual requirements, many generate incomplete models with missing elements. Furthermore, requirements traceability is largely neglected, with only two approaches addressing it and lacking explicit detail on how traceability links are maintained during the transformation process.
This review emphasize the need for formalized evaluation techniques and greater transparency and accessibility of approaches used in the early-stage requirements transformation.
 
\end{abstract}

\keywords{Textual Requirements \and Analysis Models \and Model Transformation}

\section{Introduction}

The requirement analysis phase in an any software development project is crucial, laying the foundation for the success or failure of the project. 
It is of paramount importance to ensure that analysis models generated during the requirements analysis phase are a true representation of requirements in their natural form. Analysis models play a crucial role by serving as the blueprint for design, development, and eventual implementation; faults not detected at this stage can be costly to fix later \cite{Letsholo13}. However, transforming early-stage requirements into actionable plans poses significant challenges, often due to their inherent ambiguity, complexity, and variability. Over the past decades, numerous approaches and methodologies have been proposed to tackle these challenges, each bringing unique perspectives and techniques to the table.

The need for automation was realized in the early 1990s \cite{Derr95}, and the trend in research in the area of automation of requirements analysis indicates that this area accrued interest from the early 2000s \cite{Yue11}. This is evident from the publication of journal and conference articles in this area from the early 2000s. According to Bhala et al. \cite{Vidya14}, there are two main reasons for this renewed interest in the field. First, subsequent development in automatic text analysis, like parsers for natural language, tagged corpora for specific needs and anaphora resolvers, which have reduced the complexities of dealing with natural language. Second, the evolutions of diagramming models like UML, and the ongoing attempts at standardising contents of models. With the software industry converging to a few standard and well-known formats of requirements documentation, more focused research with wider application is possible.

Yue et al. \cite{Yue11} conducted the first extensive survey of automation approaches in the area of requirements analysis in 2011. Yue's survey investigated transformation approaches between user requirements and analysis models. The review led to the analysis of 20 primary studies (16 approaches) obtained after a carefully designed procedure for selecting papers published in journals and conferences from 1996 to 2008, and Software Engineering textbooks. Seven of the approaches were automated, and only four of the automated works could generate class diagrams, object diagrams or conceptual models, while the rest generated other types of analysis models like sequence diagrams or state charts. The resulting models from the transformation processes of the papers surveyed, were found to have low efficiency, and lacked evaluation mechanisms. 

This systematic review aims to investigate, analyze, and synthesize the plethora of early-stage requirements transformation approaches documented in the literature from 2000 to 2014. By critically examining these approaches, we seek to identify trends, best practices, and gaps within current methodologies. Through this comprehensive analysis, stakeholders—including researchers, practitioners, and project managers—can gain deeper insights into effective strategies for managing and transforming initial requirements, ultimately improving the quality and efficiency of software development processes.

In the subsequent sections, we outline the review methodology, provide an overview of early-stage requirements transformation approaches, delve into detailed analyzes of these approaches, and discuss the findings. 

\section{Methodology}

The main objective for the review methodology is to construct a conceptual framework on which existing work can be synthesised and compared. This review follows the guidelines for performing systematic literature reviews in software engineering presented in \cite{Kitchenham07}. Furthermore, the review builds on the systematic review of transformation approaches between user requirements and analysis models conducted by Yue et al. \cite{Yue11}. Building on Yue et al.'s study, this review investigates progress made from 2000 to 2014 and identifies prevailing issues that still need to be addressed.
The scope of my review is limited to transformation approaches that take as input, requirements specified in natural language (i.e., structured or unstructured). The resultant output or target model, however, can represent either the structural (e.g., class model) or behavioural (e.g., sequence diagram) aspects of the system being developed.

This review was guided by the following five research questions: 

\begin{enumerate}
    \item[RQ1] Are there any intermediate models used during the transformation of requirements into analysis models? If any, how do they affect the efficiency of the transformation?
    \item[RQ2] What are the types of analysis models produced by these transformations? 
    \item[RQ3] Do these approaches include traceability management support? If yes, what techniques are used for establishing and maintaining traceability links between a requirements specification and its analysis model? 
    \item[RQ4] Are these approaches automated, automatable, semi-automated or manual?
    \item[RQ5] What evaluation methods have been applied to evaluate these approaches?  
\end{enumerate}

\subsection{Search Strategy}
The search strategy was primarily directed at finding published articles from the year 2000 to February 2014---in the contents of ACM Digital library, IEEE Explore, ScienceDirect, SpringerLink and Google Scholar. Primary studies were selected from articles published in peer-reviewed journals in software engineering literature, which are \textit{IEEE Transaction on Software Engineering} (TSE); \textit{ACM Transactions on Software Engineering and Methodology} (TOSEM); \textit{Information and Software Technology} (IST); \textit{Journal of Systems and Software} (JSS). More studies were selected from top conferences in software engineering, such as, \textit{International Conference on Software Engineering} (ICSE); \textit{Automated Software Engineering} (ASE); \textit{Foundations on Software Engineering} (FSE); \textit{Requirements Engineering} (RE) and \textit{International Conference on Model Driven Engineering Languages and Systems} (MODELS). 
To ensure that relevant articles are not missed, additional searches were performed directly on websites for key journals and conference proceedings. Additionally, to augment the collection of primary studies, the reference lists of identified primary studies were scanned to identify additional papers. 

The search was initiated by the construction of a search string to perform electronic searches. The search string was based on the research questions. There are three groups of search terms used: population terms, intervention terms and outcome terms. The search string was formed by using a disjunction of the keywords of each term group and the conjunction of the three groups.
The term groups are borrowed from the medical guidelines for considering a question about the effectiveness of a treatment \cite{Kitchenham07}. These groups are discussed below in the context of software engineering, and relevant terms forming the search string are identified: 

\begin{description}
\item [Population terms:] are keywords that might represent a category of a software engineer, an application area or an industrial group. In this study, they represent the domain of transforming requirements into analysis models. The terms used include:

\{\textit{requirements analysis OR requirements engineering OR requirements refinement OR requirements formalisation OR use cases analysis OR domain modelling OR conceptual modelling OR model driven development OR model-driven development}\}

\item [Intervention terms:] are the keywords that represent a software methodology, tool, technology or procedure that address a specific issue. For this study, they represent techniques applied in the domain to achieve a transformation. The terms used include:

\{\textit{automated transformation OR automatic transformation OR model transformation OR transformation OR transform OR transforming OR translation OR translate OR translating OR derive OR deriving OR generation OR generate OR generating OR linguistic analysis OR linguistic analyze OR natural language processing}\}

\item [Outcome terms:] are keywords that relate to relevant impacts from the interventions. In this study, they represent different types of analysis models generated by transformations. The terms used include:  

\{\textit{analysis model(s) OR object model(s) OR static model(s) OR dynamic model(s) OR UML model(s) OR class diagram(s) OR sequence diagrams(s) OR interaction diagram(s) OR activity diagrams(s) OR state machine(s) OR state chart(s) OR class model(s) OR interaction model(s) OR object oriented model(s) OR object-oriented model(s) OR object(s) OR class(es) OR message sequence chart(s)}\}

\end{description}

All sources resulting  from the search were scanned to select works to be included in the review. First, a paper's title and abstract were examined to see whether it was relevant. If the title and abstract of the paper could not help make a decision, then the paper's full text was examined.

\subsection{Inclusion and Exclusion Criteria}
The electronic search using the defined search string, resulted in thousand of papers retrieved. Approaches for transforming requirements into analysis models vary a great deal as different requirement representations are adopted as inputs, and/or different analysis models are generated \cite{Yue11}.  

The review includes: 
\begin{itemize}

\item Papers published in the years 2000 to February 2014.

\item The most recent paper or the one with the most complete description of the approach, when encountering more than one paper describing the same or similar approaches,  published in different venues;

\item All papers describing different parts of a single approach, published over several papers or in different venues. 
\end{itemize} 

The review excludes:
\begin{itemize}
\item Papers not relevant to the transformation of requirements, (e.g., papers discussing information retrieval techniques from software artefacts);

\item Papers proposing transformation approaches between software artefacts that are out of the review scope, (e.g., transformations between requirements models and design models or code);

\item Papers with insufficient technical information regarding their approaches, (e.g., papers that do not provide a detailed description on transformation steps);

\end{itemize}

\subsection{Data Extraction and Synthesis}
\label{sec_cframe}

The purpose for the review criteria is to help organise and synthesis information from the identified papers in systematic and precise manner.
A starting point for every transformation is a requirements specification document.
There are several ways of documenting requirements, such as textual descriptions, use cases, customised document templates and formal specification for some critical systems. However, this review focuses on transformation approaches that specify requirements in natural language. Natural language requirements can be written using either Unstructured NL or Structured NL. 

\begin{description}
\item [Unstructured NL-based Approaches] assume that requirements are written in completely unrestricted language, and use general NL processing systems to process them. Unstructured NL refers to pure textual descriptions that do not have a pre-defined structure or model;

\item [Structured NL-based Approaches] assume requirements written in a controlled or structured language, and use ad-hoc parsing techniques or traditional context-free grammars to analyze requirements. Structured NL aims to minimise ambiguity, redundancy and complexity of requirements by controlling the grammar and vocabulary to facilitate automated analysis. 
\end{description}

The transformation approaches are first categorised into the above mentioned broad categories (i.e., Structured NL or Unstructured NL). Following this high-level classification, the approaches are then reviewed based on a set criteria, which are NL processing techniques; transformation process; intermediate models; the type of analysis models produced, and requirements traceability links.

\subsubsection{NL Processing Techniques}
NL processing techniques are needed to pre-process textual requirements before they are loaded to the transformation process. The pre-processed requirements can be outputted as intermediate models, which are further transformed into an analysis model.
There are five types of natural language processing techniques, which can be used independently or combined:
\begin{itemize}
\item \textit{Lexical Analysis} -- is concerned with the identification and processing of the basic constructs of language such as words and punctuation marks.  Lexical analysis employs techniques of tokenisation, sentence splitting, morphological analysis and part-of-speech tagging.

\item \textit{Syntactic Analysis} -- is the process of analysing a sequence of tokens to determine grammatical structure with respect to a given formal grammar. The output is usually a syntactic parse tree. 
\item \textit{Semantic Analysis} -- it is the process of adding semantic information to a parse tree to create meaningful relationships.
 
\item \textit{Categorization} -- is the process of recognising, differentiating, and classifying requirements for a specific purpose. This process is usually performed manually. 

\item \textit{Pragmatic Analysis} -- eliminates ambiguities and inconsistencies in requirements. Pragmatic analysis can be used to check the consistency of new information before adding it to an existing model.

\end{itemize}

 This review does not go into details of pre-processing techniques, it only checks whether the investigated transformation approaches use any form of requirements pre-processing, and whether this step is automated or manual.
    
\subsubsection{Intermediate Models}
Intermediate models are used in transformations to bridge the gap between requirements and the analysis model.
There are two types of intermediate models: pre-processed requirements and abstraction models. 

\begin{itemize}

\item \textit{Pre-processed Requirements} -- are the output of NL processing techniques. The output is usually a machine-readable parse tree that act as a target from NL requirements, and a source to the analysis model.

\item \textit{Abstraction Model} -- is a generic solution to solving a problem that can be applied to a wide range of specific situations. In the context of this thesis, these are the bridging abstraction between the informal world of requirements, and the more formal world of requirements modelling.  
In some cases, textual requirements have to be pre-processed first before being mapped onto abstraction models. 
\end{itemize}

This review checks for the use of intermediate models in the transformation approaches, and the impact of intermediate models in transformations.

\subsubsection{Transformation Process}
A Transformation can be either direct or indirect. A direct transformation takes NL requirements as input and translates them into an Analysis Model without any Intermediate Model used. An indirect transformation uses one or more Intermediate Models to bridge the abstraction gap between the requirements and the analysis model. Intermediate models function as temporary source or target models in transformations, that is, either from requirements to the first intermediate model, between two intermediate models and from the last intermediate model to an analysis model.
There are three types of transformations used during model construction: Rule-based, Ontology-based and Pattern-based. 

\begin{itemize}
\item \textit{Rule-based} -- transformation uses a set of predefined transformation rules. 
\item \textit{Ontology-based} -- transformation utilises shared vocabularies for describing notions of a certain domain, whose semantics is defined in a formal and machine processable form. During pre-processing these vocabularies are built into an ontology model that acts as an intermediate model in the transformation. 
\item \textit{Pattern-based} -- transformation translates source patterns into target patterns through matching and mapping processes.
\end{itemize}

This study evaluates the reviewed transformation approaches with respect to their automation capabilities. The automation criterion evaluates whether a transformation is \textit{automatic, automatable, semi-automatic} or \textit{manual}. A transformation is deemed automatic if it is fully implemented. If not implemented but the algorithm is provided then it is deemed as automatable. If the transformation requires some user intervention then it is considered semi-automatic.        

\subsubsection{Analysis Models}
An analysis model is a description of what a system is required to do functionally and aims to be less ambiguous, more correct and consistent than textual requirements \cite{Bruegge04}. An analysis model in this context represents an output of a transformation process. A complete analysis should describe two aspects of a system: structure and behaviour. 

\begin{itemize}
\item \textit{Structure} -- emphasises the things that must be present in the system being modelled using classes, objects, attributes, operations, relationships, etc. The structural aspect can be presented in a Class Diagram, Entity Relationship (ER) Model, Object Model, or a Domain Model. 

\item \textit{Behaviour} -- emphasises the dynamic behaviour of the system by showing collaborations among objects and changes to the internal states of objects. The behaviour aspect is classified into Sequence Diagram, State Chart, Activity Diagram and Process Model.   
    
\end{itemize} 

In a typical object-oriented software development process, the analysis model is usually represented as a UML model containing various diagrams and possibly constraints. This systematic review is, however, not limited to UML models. Other representations are also taken into account, as they often share similar object-oriented concepts.

 \subsubsection{Requirements Traceability Links}

Traceability is an important property in model transformations to check if the original requirements statements have been correctly transformed into the target analysis models. Three types of traceability link can be established, which are: Text2Model, Model2Model, and Inter-requirements links.
\begin{itemize}
\item \textit{Text2Model links} -- are between constructs of NL requirements and elements of intermediate models; 
\item \textit{Model2Model links} -- are between elements of an intermediate model and analysis model elements; and 
\item \textit{Inter-requirements links} -- relate a model element to several requirement statements or to different parts of an analysis model. 
\end{itemize} 

For each transformation approach evaluated, the review checks whether there is requirements traceability support built into the transformation process. If any, what are the types of traceability links supported and finally, is traceability support automated or manual? 

\subsubsection{Evaluation Methods}
It is important to ensure that the generated analysis models correctly and completely represent the problem being addressed. In addition, the transformation process has to be validated to ensure that it does what it is expected to do. The criterion checks the extent at which the transformation approaches have been evaluated, as well the method used, and obtained results. 
There are three possible evaluation methodologies, which are: Case studies, Performance Evaluation, and Human-in-the-Loop Experiments. 

\begin{itemize}

\item \textit{Case Studies} -- help evaluate the benefits of the approach in a cost effective way to ensure that the transformation process produces the expected results. Transformation approaches cannot be validated in an analytical way \cite{Yue11}, hence, extensive case studies are necessary to demonstrate the applicability of the transformation approach; 

\item \textit{Performance  Evaluation} -- is a measurement of system performance in one or more specific areas \cite{Hirschman95}. When considering methodology for measurement in a given area, a distinction is made between criterion, measure and method. \textit{Criterion}: what we are interested in evaluating, for example, correctness (i.e., how close the models produced by the system are to those produced by human analysts). \textit{Measure}: what property of system performance we report to get at the chosen criterion, for example, recall and precision metrics (i.e., ratio of hits to hits and misses). \textit{Method}: how we determine the appropriate value for a given measure and a given system, for example, benchmark comparison; and 

\item \textit{Human-in-the-loop (HITL) Experiments} -- involves humans to check the validity and completeness of the generated models. The goal is to validate that the model rightly captures the intended domain knowledge, which is a matter of mutual understanding between modelling experts and domain experts \cite{Bertolino11} (who usually are not the same people). Some experiments in this area may use people with no modelling experience, as a way of demonstrating the benefits of transformation approaches to novice modellers.  

\end{itemize}

\begin{figure}[!ht]
\centering
\includegraphics[scale=0.64]{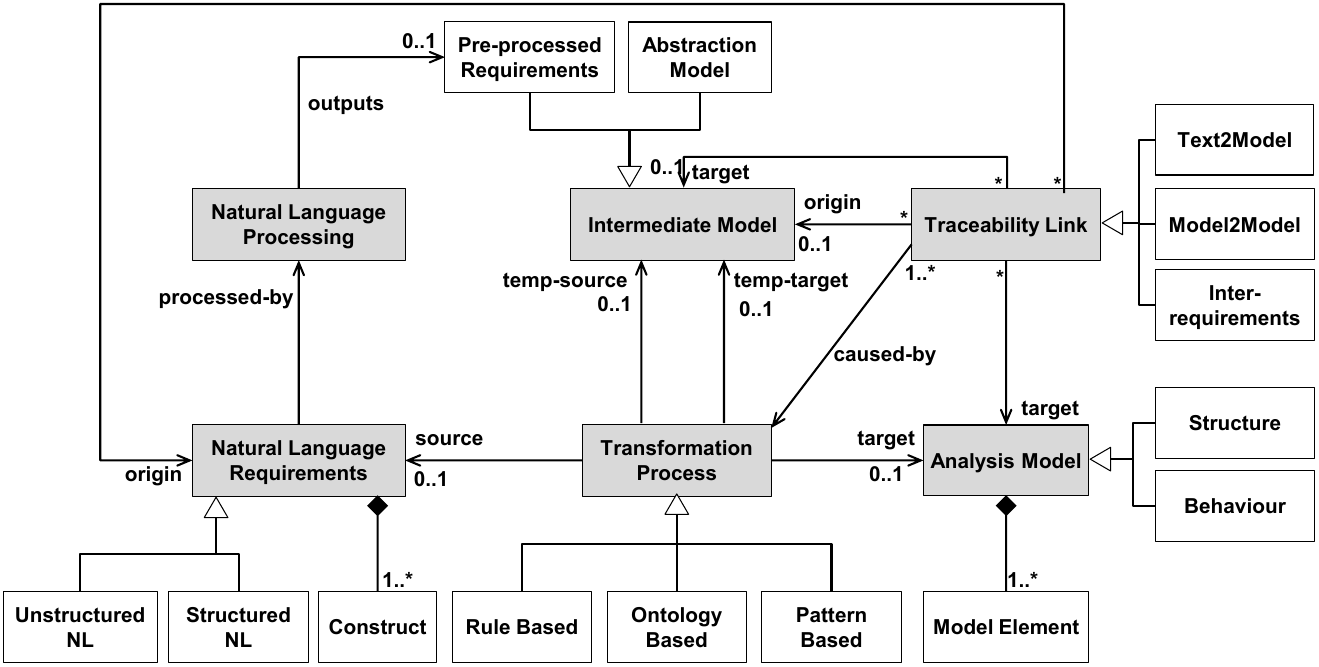}
\caption{Framework to show the relationships between various elements within the review criteria}
\label{fig:framework}
\end{figure}

The review criteria are illustrated through a conceptual framework, as shown in Figure~\ref{fig:framework}, which highlights the relationships between various elements within the review criteria.

\section{Results and Findings}

\subsection{Unstructured NL-based Approaches}
\label{sub_unstructured}

\textit{CM-Builder - Class Model Builder:} is an NL-based CASE (Computer Aided Software Engineering) tool developed by Harmain and Gaizauskas \cite{Harmain03}, \cite{Harmain00}. This tool aims to support the initial stage of the analysis process, that of, identifying object classes, attributes and relationships used to model the problem domain. The builder takes as input software requirements text written in English and constructs, either automatically or interactively with the analyst an initial UML Class Model. 
There are two versions of the tool available, CM-Builder 1 and CM-Builder 2. CM-Builder 1 generates a list of candidate classes, attributes, and relationships and assists the analyst in filtering these lists. An interactive graphical interface is provided to help the analyst build the final class model. In contrast to CM-Builder 1, there is no user interaction with CM-Builder 2: a first-cut conceptual model is produced automatically from NL requirements. The outputted model file can be loaded to a CASE tool to further refine and extend the model. 

\textit{LIDA - LInguistic assistant for Domain Analysis:} is a methodology and a prototype tool to provide linguistic assistance in the model development process \cite{Overmyer01}. The methodology involves an initial identification of model elements, or conceptualisation, through assisted text analysis, followed by a refinement through validation using text descriptions of the model being developed, in this case UML class diagram. This methodology is not fully automated; it relies heavily on the analyst's interaction. LIDA compiles a list of the words and multi-word terms in a document, and provides a graphical interface for the user to mark them as corresponding to elements of a model.
LIDA's system architecture is composed of two main components: the Text Analysing Environment and the Model Editing Environment. The Text Analysing Environment employs a domain-independent linguistic parser to analyze textual inputs such as domain descriptions and use cases. It allows analysts to mark words or phrases as candidate model elements, highlighting them according to their type (e.g., candidate classes from nouns, attributes from adjectives, methods and roles from verbs). After identifying these candidate model elements, the Model Editing Environment is used to graphically associate attributes, methods, and roles with the appropriate classes, thereby facilitating the construction of the model from the marked elements.

\textit{NL-OOPS (NL Object-Oriented Production System)} \cite{Mich02} aids in requirements analysis by generating object-oriented models from natural language (NL) requirements documents. It uses LOLITA \cite{Morgan95}, a linguistic processing system that handles various aspects of English NL requirements. LOLITA creates a Semantic Network (SemNet) as an intermediate model to bridge NL requirements with object-oriented models, addressing syntax-semantic isomorphism issues. In the first phase, LOLITA analyzes the requirements text to create the semantic network, which NL-OOPS then uses in context-independent analysis to generate a list of candidate classes. In the third phase, context-dependent analysis extracts classes, associations, attributes, and operations. The final class models can be interactively refined, with traceability functions linking back to the semantic network and original text.

\textit{CIRCE:} \cite{Ambriola06}, \cite{Ambriola03} is a system designed for the automatic analysis of natural language requirements, utilizing successive transformations to render models extracted from requirements. It generates models for the requirements document, the system described (e.g., UML class diagrams, state automata), and the requirements development process itself. CIRCE is implemented as a web-based system, where a central server acts as a repository for requirements documents. Users can edit their requirements and request specific views through a standard web browser.

\textit{Class-Gen System:} \cite{Elbendak11}, \cite{Elbendak11_thesis} is a modular NL-based CASE tool that facilitates domain-independent object-oriented analysis by partially automating the identification of objects or classes from natural language requirement specifications, specifically use case descriptions. It employs memory-based shallow parsing for syntactic analysis of English text. The tool generates a Parsed Use Case Description (PUCD) to capture parsing output, extracting nouns, verbs, adjectives, and adverbs to identify classes, attributes, and relationships. The object identification process then uses identification rules to construct an initial class model, which can be manually refined. Class-Gen's effectiveness was evaluated through performance measurement on six case studies and a comparative analysis against manually created class models by software engineer.

\textit{Ilieva \& Ormandjieva} \cite{Ilieva06} developed a methodology to aid the automatic analysis of textual requirements and the formal presentation of the extracted knowledge into requirement models, comprising three main phases. First, the Linguistic Component syntactically and semantically analyzes sentences, using grammatical rules and heuristics to label roles (e.g., subject, predicate) in compound sentences and presenting the findings in a tabular form. In the second phase, these elements are interconnected into a Semantic Network, which depicts the relationships between entities as nodes and transitions, serving as an intermediate model. Finally, the third phase translates the semantic network into object-oriented analysis model elements, including domain models, use case path models, and activity diagrams. Although there is no tool support, a simple case study illustrates the approach's phases.

\textit{Friedrich et al.} \cite{Friedrich11} proposed an approach to extract business process models directly from natural language text using syntactic and semantic analysis techniques. It utilizes a \textit{World Model} as an intermediate structure to store and organize extracted knowledge, consisting of four elements: Actor, Resource, Action, and Flow. The transformation procedure includes three steps: (1) Construct Extraction, mapping textual constructs to World Model elements; (2) Conditional Marker Detection, identifying conditional markers (e.g., "if", "then") and mapping them using control-flow patterns (ConditionIndicators, ParallelIndicators, ExceptionIndicators, SequenceIndicators); and (3) BPMN Transformation, converting the World Model into a BPMN representation by creating nodes, building sequence flows, removing dummy elements, finishing open ends, and processing meta activities.

The \textit{NIBA (Natural Language Based Requirements Analysis)} project \cite{Fliedl07}, \cite{Fliedl05} proposed a method to linguistically analyze textual specifications and translate them into a conceptual pre-design schema, which can then be mapped to representations like UML. NIBA's transformation process involves three steps:
First, Linguistic Analysis: Conducting a deep morphological, syntactic, and semantic analysis of verbal features.
Second, Schema Construction: Transforming the parsed output into a conceptual pre-design schema using the KCPM (Klagenfurt Conceptual Predesign Model), which includes key notions such as thing-types (domain concepts), connection-types (relationships), and operation-types (functional services).
Thirds, Component Mapping: Mapping the derived KCPM entries onto conceptual models (e.g., activity diagrams, state charts, class diagrams). 
While the toolset supports automatic transformation to activity diagrams, constructing state charts and class diagrams is not yet automated but can be automated.

\textit{Capuchino et al.} \cite{Capuchino00}, \cite{Juristo99} proposed a formal approach for generating object-oriented (OO) models from natural language requirements. The approach involves two main activities: conceptual model formalization and OO model creation. 
Conceptual Model Formalization: Uses rules to identify key elements by defining relationships between linguistic patterns (natural language structures) and conceptual patterns (conceptual structures), ensuring their correspondence through predicate logic and set theory.
OO Model Creation: Utilizes the formalization results to guide analysts in building two OO models: the object model (static structure) and the behavior model (dynamic aspect). The approach also includes a validation process using familiar natural language sentences, facilitating easier validation compared to abstract conceptual models.

\textit{Shinde et al.} \cite{Shinde12} proposed a rule-based system for NL-based object oriented software modelling. The proposed system uses NL processing techniques to lexically analyze software requirement texts written in English and builds an integrated discourse model of the processed text, represented in a Semantic Network. The system automatically constructs UML diagrams (i.e., class model representing the object classes mentioned in the text and their relationships) from the Semantic Network. The system uses WordNet dictionary to get the semantic relations between the classes and attributes. 
In addition, the system includes an interactive user interface that manages tasks such as creating, printing, saving and analysing requirements. It also handles the graphical representation of the diagrams and allows the user to add, delete, rename classes and relationships in the class diagram. Through this interface, the domain expert can help improve the results obtained by the system.

\textit{RACE - Requirements Analysis \& Class Diagram Extraction} \cite{Ibrahim10} is a method and tool to facilitate requirements analysis process and class diagram extraction from textual requirements using natural language processing and domain ontology techniques. RACE tool uses an OpenNLP\footnote{OpenNLP ---http://opennlp.apache.org/} parser to obtain lexical and syntactic parses of textual requirements written in English. Furthermore, it uses WordNet to validate the semantic correctness of the sentences generated during syntactic analysis. Concepts are identified according to the requirements document---domain ontology is used to improve the performance of concepts identification. RACE tool is able to find concepts based on nouns, noun phrases and verbs analysis. Heuristic rules are then used to extract class diagram from the identified concepts.
The RACE tool assists analysts by providing an efficient and fast way to produce the class diagram from their requirements. It supports a good interaction with users by providing a human-centred user interface which makes the user a part of the analysis process \cite{Ibrahim10}.

\textit{Li et al.} \cite{Li04} proposed an algorithmic approach to analysing and restructuring natural language text. This approach is a tentative start
and focuses on analysing textual requirements in order to elicit object-oriented system elements, such as classes, objects, methods, properties and relationships. The algorithm supports the analysis process, and rule set for mapping NL and object-oriented elements, and the standardization of NL structure. The process includes the following steps. Step 1: POS tagging on NL-based domain description, which identifies the nouns, verbs and pronouns. Step 2: the POS identification is simplified in standard sentence structure (\textit{subject--verb--object}) based on the NL structure standardization rules. Step 3: the initial class diagram is generated accordingly from the simplified NL description. Users are involved in Step 4 to confirm the classes implied by pronouns and actors.

Table~\ref{tab_unstructured} presents a summary of 12 unstructured NL-based approaches reviewed. An in-depth analysis of the table is presented in Section~\ref{sec_synthesis}.
    
\begin{table}[!ht]
\renewcommand{\arraystretch}{1.2}
\centering
\footnotesize
\begin{tabular}{cc|r|c|c|c|c|c|c|c|c|c|c|c|c|} \\ \cline{4-15} 
&\multicolumn{1}{l}{} &\multicolumn{1}{l|}{} &  \multicolumn{12}{c|}{\textbf{Approaches}}  \\ \cline{4-15}
 
\multicolumn{3}{c|}

& {\rotatebox[origin=c]{90}{CM-Builder}} 
& {\rotatebox[origin=c]{90}{LIDA}}
& {\rotatebox[origin=c]{90}{NL-OOPS}}
& {\rotatebox[origin=c]{90}{CIRCE}}
& {\rotatebox[origin=c]{90}{Class-Gen}}
& {\rotatebox[origin=c]{90}{Ilieva et al.}}
& {\rotatebox[origin=c]{90}{Friedrich et al.}}
& {\rotatebox[origin=c]{90}{NIBA}} 
& {\rotatebox[origin=c]{90}{Capuchino et al.}} 
& {\rotatebox[origin=c]{90}{Shinde et al.}} 
& {\rotatebox[origin=c]{90}{RACE}} 
& {\rotatebox[origin=c]{90}{Li et al.}}\\ \hline \hline
 
\multicolumn{1}{|l}{\multirow{5}{*}{\rotatebox[origin=c]{90}{\textbf{NL}}}} &\multicolumn{1}{l|}{\multirow{5}{*}{\rotatebox[origin=c]{90}{\textbf{Processing}}}}
			          &Lexical  &\ding{51} &\ding{51} &\ding{51} &\ding{51} & & \ding{51}& &\ding{51} & &\ding{51} &\ding{51} &\ding{51}\\ \cline{3-15}    
\multicolumn{2}{|l|}{}&Syntactic&\ding{51} &\ding{51} &\ding{51} &\ding{51} &\ding{51} & &\ding{51} &\ding{51} & & &\ding{51} &\\ \cline{3-15}
\multicolumn{2}{|l|}{}&Semantic &\ding{51} & &\ding{51} &\ding{51} & &\ding{51} &\ding{51} &\ding{51} & & & &\\ \cline{3-15}
\multicolumn{2}{|l|}{}&Categorisation & & & & & & & & &\ding{51} & & & \\ \cline{3-15}
\multicolumn{2}{|l|}{}&Pragmatic  & & &\ding{51} & & & & & & & & &\\ \hline\hline

\multicolumn{2}{|l|}{\multirow{2}{*}{\textbf{IM}}}
			          &Preprocessed NL& & & &\ding{51} & & & & & & & &\\ \cline{3-15}    
\multicolumn{2}{|l|}{}&Abstraction Model&\ding{51} & &\ding{51} & &\ding{51} &\ding{51} &\ding{51} &\ding{51} &\ding{51} &\ding{51} & &\\ \hline\hline

\multicolumn{1}{|l|}{\multirow{7}{*}{\rotatebox[origin=c]{90}{\textbf{Transformation}}}}
  &{\multirow{3}{*}{\rotatebox[origin=c]{90}{Process}}}
			          &Rule-based  &\ding{51} &\ding{51} &\ding{51} &\ding{51} &\ding{51} &\ding{51} &\ding{51} &\ding{51} &\ding{51} &\ding{51} &\ding{51} &\ding{51}\\ \cline{3-15}    
\multicolumn{1}{|l|}{}&\multicolumn{1}{l|}{}&Ontology-based&\ding{51} & & & & & & & & & &\ding{51} &\\ \cline{3-15}
\multicolumn{1}{|l|}{}&\multicolumn{1}{l|}{}&Pattern-based & & & & & & & & & & & &\\ \cline{2-15}
\multicolumn{1}{|l|}{}&{\multirow{4}{*}{\rotatebox[origin=c]{90}{Automation}}}
  &Automatic&\ding{51} & &\ding{51} &\ding{51} & & &\ding{51} & & & &\ding{51} &\\ \cline{3-15}
\multicolumn{1}{|l|}{}&\multicolumn{1}{l|}{}&Semi-Auto& &\ding{51} & & &\ding{51} & & &\ding{51} & &\ding{51} & &\\ \cline{3-15}
\multicolumn{1}{|l|}{}&\multicolumn{1}{l|}{}&Automatable& & & & & &\ding{51} & & & & & &\\ \cline{3-15}
\multicolumn{1}{|l|}{}&\multicolumn{1}{l|}{}&Manual& & & & & & & & &\ding{51} & & &\ding{51}\\ \hline\hline

\multicolumn{1}{|l|}{\multirow{8}{*}{\rotatebox[origin=c]{90}{\textbf{Analysis Models}}}}
&{\multirow{4}{*}{\rotatebox[origin=c]{90}{Structure}}}
					  &Class Model  &\ding{51} &\ding{51} &\ding{51} &\ding{51} &\ding{51} & & &\ding{51} &\ding{51} &\ding{51} &\ding{51} &\ding{51}\\ \cline{3-15} 
\multicolumn{1}{|l|}{}&\multicolumn{1}{l|}{}&ER Diagram  & & & &\ding{51} & & & & & & & &\\ \cline{3-15} 
\multicolumn{1}{|l|}{}&\multicolumn{1}{l|}{}&Domain Model& & & & & &\ding{51} & & & & & &\\ \cline{3-15} 
\multicolumn{1}{|l|}{}&\multicolumn{1}{l|}{}&Use Cases   & & & & & &\ding{51} & & & & & &\\ \cline{2-15}

\multicolumn{1}{|l|}{}&{\multirow{4}{*}{\rotatebox[origin=c]{90}{Behaviour}}}
  &State Chart& & & &\ding{51} & & & &\ding{51} & & & &\\ \cline{3-15}
\multicolumn{1}{|l|}{}&\multicolumn{1}{l|}{}&Activity Dia.& & & & & &\ding{51} & &\ding{51} &\ding{51} & & &\\ \cline{3-15}
\multicolumn{1}{|l|}{}&\multicolumn{1}{l|}{}&Sequence Dia.& & & & & & & & & & & &\\ \cline{3-15}
\multicolumn{1}{|l|}{}&\multicolumn{1}{l|}{}&BPMN  & & & & & & &\ding{51} & & & & &\\ \hline\hline

\multicolumn{2}{|l|}{\multirow{3}{*}{\textbf{RTL}}}
			          &Text2Model  & & &\ding{51} & & & & & & & & &\\ \cline{3-15}    
\multicolumn{2}{|l|}{}&Model2Model & & &\ding{51} & & & & & & & & &\\ \cline{3-15}
\multicolumn{2}{|l|}{}&Inter-Requirements  & & & & & & & & & & & &\\ \hline\hline

\multicolumn{2}{|l|}{\multirow{3}{*}{\textbf{EM}}}
			          &Case Study  &\ding{51} &\ding{51} &\ding{51} &\ding{51} &\ding{51} &\ding{51} & &\ding{51} &\ding{51} &\ding{51} &\ding{51} &\ding{51}\\ \cline{3-15}    
\multicolumn{2}{|l|}{}&HITL& & & & &\ding{51} & & & &\ding{51} & & &\\ \cline{3-15}
\multicolumn{2}{|l|}{}&Performance Eval. &\ding{51} & & & &\ding{51} & &\ding{51} & & & & &\\ \hline\hline

\multicolumn{15}{l}{IM - Intermediate Models; RTL - Requirements Traceability Links;} \\ 					
\multicolumn{15}{l}{EM - Evaluation Methods; HITL - Human-in-the-Loop Experiments.} \\  

\end{tabular}

\caption{A Summary of Unstructured NL-based Approaches}
\label{tab_unstructured}
\end{table}

\subsection{Structured NL-based Approaches}
\label{sub_structured}

\textit{Bhala et al.} \cite{Vidya14} developed an automated system to create conceptual models from grammatically correct natural language functional specifications, excluding negative statements and non-functional requirements, with each sentence being self-contained. The process involves four functional modules: text preprocessing, which performs natural language processing tasks like word tokenization and stemming; syntactic feature extraction, which uses syntactic analysis to extract grammatical constructs as typed dependencies; design elements extraction, which examines these dependencies against rules to create classes, attributes, objects, or relationships; and relation type classification, which categorizes the extracted relationships as associations, aggregation, composition, or inheritance. The approach successfully generates conceptual models with high precision and recall, evaluated against both expert and novice human judgments. 

\textit{UCDA - Use-Case driven Development Assistant} \cite{Liu04} automates the development of use case diagrams, writing of use case specification documents, and creation of class models from natural language requirements using a natural language parser and Rational Rose's extensibility interface. The process begins with stakeholders' requests, which are parsed into sentences and abstract syntactic structures. These are analyzed to identify actors and use cases, forming a use case diagram. Detailed use case information is gathered from stakeholders, formatted, validated, and compiled into a use case specification document. This document, along with the use case diagram, serves as the basis for automated class model generation using predefined rules. While not all activities are fully automated—such as use-case specification—the user interacts with UCDA to provide necessary information. Generated diagrams can be visualized in Rational Rose.

\textit{UMGAR - UML Model Generator} \cite{Deeptimahanti11}, \cite{Deeptimahanti09} is a domain-independent semi-automated tool that aids developers in generating UML models from natural language (NL) requirements. The tool's process architecture comprises two main components: normalizing requirements and the model generator. The normalizing requirements component removes ambiguities and identifies incomplete requirements, using syntactic reconstruction rules to simplify complex requirements into active voice sentences. The model generator component generates various OO models like use-case diagrams, analysis class models, and collaboration diagrams from these normalized requirements. UMGAR uses specific rules to identify candidate classes from noun phrases, attach attributes and methods, and eliminate redundancies and ambiguities. While traceability support is not detailed, it is claimed to be provided via Key-Word-In-Concept and concept location features. UMGAR includes a generic XMI parser for visualizing generated models in any UML modeling tool.

\textit{OOExpert Framework} \cite{Wahono02} is a methodology for object identification and refinement from software requirements, aimed at simplifying the understanding of the object-oriented paradigm and aiding modellers in identifying and refining objects. The methodology involves end-users actively specifying requirements using Object-Based Formal Specification (OBFS), a semi-formal template that addresses ambiguity, incompleteness, and inconsistency in requirements. The object-model creation process starts with identifying relevant objects and their associations from the domain (typically nouns in requirements statements), followed by eliminating irrelevant objects using rule-based reasoning. The attributes and behavioral properties of objects are then identified to create a class model. This methodology, demonstrated through an air traffic control project example, is supported by a system called OOExpert.

\textit{Metamorphosis Approach}, developed by D\'{i}az et al. \cite{Diaz05}, is an approach designed to model interactions and establish persistent relationships between functional requirements (Use Case Model) and their analysis (Interaction Model). Interaction granules can vary: a \textit{fragment} from an action, an \textit{interaction unit} from a use case, or an \textit{interaction model} from Use Case Models. The approach uses roles to capture linguistic abstractions and describe interactions, detailing the semantic properties of participants. The central feature is action transformation using generic, domain-independent, and implementation-independent patterns, which describe how to transform semantic contexts into semantic fragments and turn context participants into fragment elements. Combining these fragments from use case sentences results in a complete interaction diagram.

\textit{UCEd - Use Case Editor} \cite{Some06}, \cite{Samarasinghe05} automates the extraction of domain information from requirements by identifying various types of domain model entities: \textit{Concept} (important entity types/classes), \textit{System Concept} (the system as a black box), \textit{Aggregation} (“part-of” relationships), \textit{Attribute} (features of concepts), and \textit{Object} (instances of concepts). It uses a restricted natural language for use case descriptions and relies on syntactical analysis and a set of mapping rules to identify domain entities. When constructs cannot be automatically mapped, UCEd prompts the user for input, offering a semi-automated extraction mechanism due to the inherent ambiguity of natural language.

Table~\ref{structured} presents a summary of six structured NL-based approaches reviewed. An in-depth analysis of the table is presented in Section~\ref{sec_synthesis}.

\begin{table} [!ht]
\renewcommand{\arraystretch}{1.2}
\centering
\footnotesize
\begin{tabular}{cc|r|c|c|c|c|c|c|} \\ \cline{4-9} 
 &\multicolumn{1}{c}{} &\multicolumn{1}{c|}{} &\multicolumn{6}{c|}{\textbf{Approaches}}  \\ \cline{4-9}
 
\multicolumn{2}{c}

&
& {\rotatebox[origin=c]{90}{Bhala et al.}} 
& {\rotatebox[origin=c]{90}{UCDA}}
& {\rotatebox[origin=c]{90}{UMGAR}}
& {\rotatebox[origin=c]{90}{OOExpert}}
& {\rotatebox[origin=c]{90}{Metamorphosis}}
& {\rotatebox[origin=c]{90}{UCEd}}\\ \hline \hline
 
\multicolumn{1}{|l}{\multirow{5}{*}{\rotatebox[origin=c]{90}{\textbf{NL}}}} &\multicolumn{1}{l|}{\multirow{5}{*}{\rotatebox[origin=c]{90}{\textbf{Processing}}}}
			          &Lexical  & &\ding{51} & &\ding{51} & & \\\cline{3-9}    
\multicolumn{2}{|l|}{}&Syntactic&\ding{51} & &\ding{51} & &\ding{51} &\ding{51} \\\cline{3-9}
\multicolumn{2}{|l|}{}&Semantic & & & & &\ding{51} & \\\cline{3-9}
\multicolumn{2}{|l|}{}&Categorisation & & & &\ding{51} & & \\\cline{3-9}
\multicolumn{2}{|l|}{}&Pragmatic  & & & & & & \\ \hline\hline

\multicolumn{2}{|l|}{\multirow{2}{*}{\textbf{IM}}}
			          &Preprocessed NL& & & & & & \\\cline{3-9}    
\multicolumn{2}{|l|}{}&Abstraction Pattern& & & &\ding{51} &\ding{51} & \\ \hline\hline

\multicolumn{1}{|l|}{\multirow{7}{*}{\rotatebox[origin=c]{90}{\textbf{Transformation}}}}
  &{\multirow{3}{*}{\rotatebox[origin=c]{90}{Process}}}
			          &Rule-based  &\ding{51} &\ding{51} &\ding{51} &\ding{51} &\ding{51} &\ding{51} \\ \cline{3-9}    
\multicolumn{1}{|l|}{}&\multicolumn{1}{l|}{}&Ontology-based& & & & & & \\ \cline{3-9}
\multicolumn{1}{|l|}{}&\multicolumn{1}{l|}{}&Pattern-based  & & & & &\ding{51} & \\ \cline{2-9}

\multicolumn{1}{|l|}{}&{\multirow{4}{*}{\rotatebox[origin=c]{90}{Automation}}}
				&Automatic&\ding{51} &\ding{51} & & & \ding{51}& \\ \cline{3-9}
\multicolumn{1}{|l|}{}&\multicolumn{1}{l|}{}&Semi-Auto& & &\ding{51} &\ding{51} & &\ding{51} \\ \cline{3-9}
\multicolumn{1}{|l|}{}&\multicolumn{1}{l|}{}&Automatable& & & & & & \\ \cline{3-9}
\multicolumn{1}{|l|}{}&\multicolumn{1}{l|}{}&Manual& & & & & & \\ \hline\hline

\multicolumn{1}{|l|}{\multirow{8}{*}{\rotatebox[origin=c]{90}{\textbf{Analysis Models}}}}
&{\multirow{4}{*}{\rotatebox[origin=c]{90}{Structure}}}
		&Class Model  &\ding{51} &\ding{51} &\ding{51} &\ding{51} & & \\ \cline{3-9} 
\multicolumn{1}{|l|}{}&\multicolumn{1}{l|}{}&Collaboration Dia. & & &\ding{51} & & & \\ \cline{3-9} 
\multicolumn{1}{|l|}{}&\multicolumn{1}{l|}{}&Domain Model  & & & & & &\ding{51} \\ \cline{3-9} 
\multicolumn{1}{|l|}{}&\multicolumn{1}{l|}{}&Use Cases  & & &\ding{51} & & & \\ \cline{2-9}

\multicolumn{1}{|l|}{}&{\multirow{4}{*}{\rotatebox[origin=c]{90}{Behaviour}}}
           &State Chart& & & & & & \\ \cline{3-9}
\multicolumn{1}{|l|}{}&\multicolumn{1}{l|}{}&Activity Dia.  & & & & & & \\ \cline{3-9}
\multicolumn{1}{|l|}{}&\multicolumn{1}{l|}{}&BPMN  & & & & & & \\ \cline{3-9}
\multicolumn{1}{|l|}{}&\multicolumn{1}{l|}{}&Sequence Dia.  & & & & &\ding{51} & \\ \hline\hline

\multicolumn{2}{|l|}{\multirow{3}{*}{\textbf{RTL}}}
			          &Text2Model  & & &\ding{51} & & & \\\cline{3-9}    
\multicolumn{2}{|l|}{}&Model2Model& & & & & & \\\cline{3-9}
\multicolumn{2}{|l|}{}&Inter-Requirements  & & & & & & \\ \hline\hline

\multicolumn{2}{|l|}{\multirow{3}{*}{\textbf{EM}}}
			          &Case Study  &\ding{51} &\ding{51} &\ding{51} &\ding{51} &\ding{51} &\ding{51} \\\cline{3-9}    
\multicolumn{2}{|l|}{}&HITL&\ding{51} & & & & & \\ \cline{3-9}
\multicolumn{2}{|l|}{}&Performance Eval. &\ding{51} & & & & & \\ \hline\hline
					  
\multicolumn{8}{l}{IM - Intermediate Models;} \\ 					
\multicolumn{8}{l}{RTL - Requirements Traceability Links;} \\ 
\multicolumn{8}{l}{EM - Evaluation Methods;} \\ 
\multicolumn{8}{l}{HITL - Human-in-the-Loop Experiments.} \\
\end{tabular}

\caption{A Summary of Structured NL-based Approaches}
\label{structured}
\end{table}

\section{Discussion}
\label{sec_synthesis}

This discussion compares approaches on the basis of the review criteria presented in Section~\ref{sec_cframe}) in order to abstract away from minor differences or detail, and to allow for the analysis of general patterns observed. The review analyzed 25 studies, categorised into 18 transformational approaches. Eleven out of 18 approaches investigated were also analyzed by Yue et al. \cite{Yue11}. Seven approaches,  Bhala et al., Class-Gen, UMGAR, Friedrich et al., Shinde et al., RACE, and Li et al. were recently published, thus, they are not part of Yue et al.'s survey. Seven approaches were published in peer-reviewed journals Capuchino et al., CM-Builder, Bhala et al., NIBA, CIRCE, Class-Gen, and UCEd; other approaches were published in conference proceedings.    

\textbf{NL Processing Techniques.} 
As shown in Table~\ref{tab_unstructured} \&~\ref{structured}, all the approaches apply at least one of the NL processing techniques. Unstructured NL-based approaches require a combination of pre-processing techniques, while structured NL-based approaches are capable of using only one technique.
The commonly used NL processing system is the Stanford parser \cite{klein03}.  The parser is capable of performing both syntactic and semantic analysis. That is, assigning part-of-speech tags to the words in the text (e.g., noun, verb, adjective) and creating grammatical structure of sentences, thus, adding meaning to the text.

During pre-processing of requirements, an external resource such as the WordNet lexical database \cite{Miller90}, \cite{Fellbaum10} is used for word sense disambiguation and morphological analysis by most approaches (e.g., CM-Builder, NL-OOPS, LIDA, UMGAR, Class-Gen, RACE). Word sense disambiguation is an open problem in NL processing which governs the process of identifying the meaning of words in context, when words have multiple meanings.

\textbf{Intermediate Models.} 
As shown in Table~\ref{tab_unstructured} \&~\ref{structured}, 11 approaches use intermediate models when a direct transformation cannot be achieved from requirements to an analysis model. Transformation rules of these indirect transformation approaches contain two set of rules: transformation rules from requirements to intermediate models, and transformation rules from intermediate models to an analysis model. The intermediate models act as the target models of the first rule set and the source models of the second rule set.
It worth noting that approaches that use intermediate models in their transformation process are capable of generating both structural and behavioural models. This is because intermediate models are abstract, domain-independent, and language-independent, making it possible to map to any modelling language.
The following considerations should be taken into account when selecting an intermediate model for an automated transformation \cite{Yue11}:

\begin{itemize}
\item The representation of the source and target models, as well as transformation rules;
\item Whether it is easy to integrate the intermediate model(s) into existing tool support;
\item If user interventions are required during transformations, it is important that users easily understand an intermediate model;
\item Whether the intermediate model is general enough for multiple purpose usage, such as generating not only the structural aspect of a system but also the behavioural aspect;
\item Whether it is directly applied to textual requirements, or to pre-processed requirements;
\item Whether it is capable of supporting requirements traceability management.
\end{itemize}

\textbf{Transformation Process.}
Eight out of 18 approaches are automatic; seven require user intervention to perform the transformation (semi-auto); two approaches are manual, and one is automatic. Generally, automatic transformations require complex pre-processing techniques, except for the UCDA approach, which only requires lexical analysis, mainly because it uses structured text as input. Rule-based transformation is the most frequently used in the model creation process. Two approaches, CM-Builder and RACE, use both an ontology and rules in their transformation process; and one approach, Metamorphosis, uses pattern-based transformation.

\textbf{Analysis Models.} 
Sixteen approaches out of 18 can derive structural models (e.g., class model, domain model, use cases) from NL requirements. A class model is used as a target model in 14 approaches. UML is the most frequently used language to represent target analysis models. Six approaches are capable of generating behavioural models (e.g., state chart, activity diagram), one of which is a process model (e.g., BPMN) generated by Friedrich et al. approach. 
A methodological open issue identified in this review is that many of the approaches  generate incomplete analysis models. Incomplete in the sense that transformation approaches are not capable of deriving from textual requirements some of the important elements of the analysis model. For example, most approaches do not derive important class model elements such as dependencies, aggregations, and cardinalities, as shown in Table~\ref{classelements}. In fact, none of the approaches represent dependency relationships in their generated class models. 

\begin{table}[!ht]
\centering
\scriptsize
\renewcommand{\arraystretch}{1.2}
\begin{tabular}{cl|c|c|c|c|c|c|c|c|c|} \cline{3-11}

&  &\multicolumn{9}{c|}{\textbf{Class Model Element}} \\ \cline{3-11}
&  & Class & Operation& Attribute& Association& Aggregation& Composition& Generalisation& Dependency &Multiplicity\\ \hline

\multicolumn{1}{|l|}{ \parbox[t]{2mm}{\multirow{12}{*}{\rotatebox[origin=c]{90}{\textbf{\textbf{Approaches}}}}}} 
    
& CM-Builder	&\ding{51}&\ding{55}&\ding{51}&\ding{51}&\ding{51}&\ding{51}&\ding{55}&\ding{55}&\ding{51} \\ \cline{2-11}

\multicolumn{1}{|l|}{} & Bhala et al.&\ding{51}&\ding{51}&\ding{51}&\ding{51}&\ding{51}&\ding{51}&\ding{51}&\ding{55}&\ding{51} \\ \cline{2-11}

\multicolumn{1}{|l|}{}& LIDA&\ding{51}&\ding{51}&\ding{51}&\ding{55}&\ding{55}&\ding{55}&\ding{55}&\ding{55}&\ding{55} \\ \cline{2-11}

\multicolumn{1}{|l|}{}& UCDA&\ding{51}&\ding{51}&\ding{55}&\ding{51}&\ding{55}&\ding{55}&\ding{55}&\ding{55}&\ding{55} \\ \cline{2-11}

\multicolumn{1}{|l|}{}& NL-OOPS&\ding{51}&\ding{51}&\ding{51}&\ding{51}&\ding{51}&\ding{51}&\ding{55}&\ding{55}&\ding{55} \\ \cline{2-11}

\multicolumn{1}{|l|}{}& CIRCE&\ding{51}&\ding{51}&\ding{51}&\ding{51}&\ding{51}&\ding{51}&\ding{55}&\ding{55}&\ding{55} \\ \cline{2-11}

\multicolumn{1}{|l|}{}& Class-Gen&\ding{51}&\ding{51}&\ding{51}&\ding{51}&\ding{51}&\ding{51}&\ding{55}&\ding{55}&\ding{51} \\ \cline{2-11}

\multicolumn{1}{|l|}{}& UMGAR&\ding{51}&\ding{51}&\ding{55}&\ding{51}&\ding{55}&\ding{55}&\ding{51}&\ding{55}&\ding{55} \\ \cline{2-11}

\multicolumn{1}{|l|}{}& Capuchino et al.&\ding{51}&\ding{55}&\ding{51}&\ding{51}&\ding{51}&\ding{51}&\ding{51}&\ding{55}&\ding{51} \\ \cline{2-11}

\multicolumn{1}{|l|}{}& Shinde et al.&\ding{51}&\ding{51}&\ding{51}&\ding{51}&\ding{51}&\ding{51}&\ding{51}&\ding{55}&\ding{55} \\ \cline{2-11}

\multicolumn{1}{|l|}{}& RACE&\ding{51}&\ding{55}&\ding{51}&\ding{51}&\ding{51}&\ding{51}&\ding{51}&\ding{55}&\ding{55} \\ \cline{2-11}

\multicolumn{1}{|l|}{}& Li et al.&\ding{51}&\ding{51}&\ding{51}&\ding{55}&\ding{55}&\ding{51}&\ding{55}&\ding{55}&\ding{55} \\ \hline\hline

\multicolumn{1}{l}{}& &12 &9 &10 &10 &8 &9 &5 &0 &4 \\ \cline{3-11}

\end{tabular}
\caption{Class diagram elements derived by the approaches.} 
\label{classelements}
\end{table}

\textbf{Requirements Traceability Links.}
Only two transformation approaches, NL-OOPS and UMGAR report on traceability. NL-OOPS claims to provide traceability functions for establishing links between elements in the analysis model, nodes in the intermediate models, and constructs of the original text. On the other hand, UMGAR claims to provide traceability between requirements and analysis models by associating keywords in requirements to model elements. However, both approaches do not explicitly discuss how traceability links are established and maintained through the transformation process.   

Deriving traceability links from transformation approaches that do not involve intermediate models should be straightforward. However, for those that require one or more intermediate models, it is a complex, yet an indispensable step since from the users' perspective it is very important to access derived traceability links between requirements and analysis models without having to deal with the intermediate model(s).

Although, it is possible to provide traceability naturally by model transformations, most transformation approaches have ignored this crucial step in software development. A mechanism is required to establish and explicitly maintain traceability links between source and target of each transformation \cite{Yue11}. It is even more crucial where multiple transformation steps are involved. Traceability management is not only relevant at different levels of abstraction (e.g., NL requirements and Class Diagram), it is also essential within models at the same level of abstraction (e.g., Class Diagram and Activity Diagram). 

\textbf{Evaluation Methods.} Several approaches did not evaluate the extent to which their generated models are correct and precise. Only four approaches CM-Builder, Class-Gen, Capuchino et al., Bhala et al. out of 18 performed extensive evaluation on their transformation approaches. Most studies have used a simple case study or a running example to illustrate their approach, rather than evaluate it. The quality of an automatically generated analysis model should be evaluated by comparing it with the one manually developed by expert modellers to see how close the automated analysis model is to expert solutions. Performance evaluation method has proved to be useful in checking how the tool performs in comparison to other tools. However, a challenge in the area is that formal evaluation of transformation tools is missing, making it impossible to compare performance evaluation results of existing approaches. In addition, most transformation tools are not available for download in order to carry out independent experiments. 

\section{Conclusion}

This systematic review has highlighted important insights into the field of early-stage requirements transformation approaches. By analyzing 25 studies categorized into 18 transformational approaches, we identified both common practices and significant gaps. Notably, all approaches leverage natural language processing techniques, with tools like the Stanford parser and WordNet playing pivotal roles. Eleven approaches utilize intermediate models, underscoring their importance in bridging requirements and analysis models and facilitating the creation of both structural and behavioral models. The selection of intermediate models should consider multiple factors, including integration ease, user comprehensibility, and the ability to support requirements traceability.

The review also revealed considerable variation in transformation processes, ranging from fully automatic to semi-automatic and manual approaches, with rule-based methods being prevalent. Although a majority of the approaches can derive structural models from NL requirements, many generated incomplete analysis models lacking essential elements such as dependencies and cardinalities. Moreover, the crucial aspect of traceability was largely neglected, with only two transformation approaches addressing it without explicit detail on maintenance of traceability links.

A significant challenge identified is the lack of robust evaluation methods. Only four approaches have conducted extensive evaluations, while most relied on simple case studies or running examples. This absence of standardized evaluation methodologies, coupled with the unavailability of transformation tools for independent testing, hinders the ability to compare and verify the performance and accuracy of different approaches. This gap points to a critical need for formalized evaluation techniques and greater transparency and accessibility of tools used in the transformation process.

In conclusion, while substantial advancements have been made in early-stage requirements transformation approaches, several areas require attention to enhance their effectiveness and reliability. Emphasizing comprehensive evaluation processes, improving traceability mechanisms, and developing intermediate models that are easily integratable will significantly benefit the field. Future research should focus on addressing these identified gaps to advance the practice of transforming textual requirements into analysis models, thereby improving the overall efficiency and accuracy of model-driven development in software engineering.

\bibliographystyle{unsrt}
\bibliography{article} 

\begin{thebibliography}{10}

\bibitem{Letsholo13}
Keletso~J Letsholo, Liping Zhao, and Erol-Valeriu Chioasca.
\newblock Tram: A tool for transforming textual requirements into analysis models.
\newblock In {\em Automated Software Engineering (ASE), 2013 IEEE/ACM 28th International Conference on}, pages 738--741. IEEE, 2013.

\bibitem{Derr95}
Kurt~W. Derr.
\newblock {\em Applying OMT: A Practical Step-by-Step Guide to Using the Object Modeling Technique}.
\newblock Cambridge University Press, New York, NY, USA, 1995.

\bibitem{Yue11}
T.~Yue, L.C. Briand, and Y.~Labiche.
\newblock A systematic review of transformation approaches between user requirements and analysis models.
\newblock {\em Requirements Engineering}, 16(2):75--99, 2011.

\bibitem{Vidya14}
Vidhu Bhala~R Vidya~Sagar and S~Abirami.
\newblock Conceptual modeling of natural language functional requirements.
\newblock {\em Journal of Systems and Software}, 88:25--41, 2014.

\bibitem{Kitchenham07}
Barbara~A Kitchenham and Stuart Charters.
\newblock Guidelines for performing systematic literature reviews in software engineering, version 2.3.
\newblock {\em EBSE Technical Report}, 2007.

\bibitem{Bruegge04}
Bernd Bruegge and Allen~H Dutoit.
\newblock {\em Object-Oriented Software Engineering Using UML, Patterns and Java}.
\newblock Prentice Hall, 2 edition, 2004.

\bibitem{Hirschman95}
Lynette Hirschman and Henry~S Thompson.
\newblock Chapter 13 evaluation: Overview of evaluation in speech and natural language processing.
\newblock {\em Survey of the State of the Art in Human Language Technology}, 1995.

\bibitem{Bertolino11}
Antonia Bertolino, Guglielmo De~Angelis, Alessio Di~Sandro, and Antonino Sabetta.
\newblock Is my model right? let me ask the expert.
\newblock {\em Journal of Systems and Software}, 84(7):1089--1099, 2011.

\bibitem{Harmain03}
HM~Harmain and R~Gaizauskas.
\newblock Cm-builder: A natural language-based case tool for object-oriented analysis.
\newblock {\em Automated Software Engineering}, 10(2):157--181, 2003.

\bibitem{Harmain00}
Harmain~M Harmain and R~Gaizauskas.
\newblock Cm-builder: an automated nl-based case tool.
\newblock In {\em Automated Software Engineering, 2000. Proceedings ASE 2000. The Fifteenth IEEE International Conference on}, pages 45--53. IEEE, 2000.

\bibitem{Overmyer01}
Scott~P Overmyer, Benoit Lavoie, and Owen Rambow.
\newblock Conceptual modeling through linguistic analysis using lida.
\newblock In {\em Proceedings of the 23rd international conference on Software engineering}, pages 401--410. IEEE Computer Society, 2001.

\bibitem{Mich02}
Luisa Mich and Roberto Garigliano.
\newblock Nl-oops: A requirements analysis tool based on natural language processing.
\newblock In {\em Proceedings of Third International Conference on Data Mining Methods and Databases for Engineering, Bologna, Italy}, 2002.

\bibitem{Morgan95}
Richard Morgan, Roberto Garigliano, Paul Callaghan, Sanjay Poria, Mark Smith, and Chris Cooper.
\newblock University of durham: description of the lolita system as used in muc-6.
\newblock In {\em Proceedings of the 6th conference on Message understanding}, pages 71--85. Association for Computational Linguistics, 1995.

\bibitem{Ambriola06}
Vincenzo Ambriola and Vincenzo Gervasi.
\newblock {On the Systematic Analysis of Natural Language Requirements with CIRCE}.
\newblock {\em Automated Software Engineering}, 13(1):107--167, 2006.

\bibitem{Ambriola03}
V~Ambriola and V~Gervasi.
\newblock The circe approach to the systematic analysis of nl requirements.
\newblock {\em Universita di Pisa, Tech. Rep. TR-03-05}, 2003.

\bibitem{Elbendak11}
Mosa Elbendak, Paul Vickers, and Nick Rossiter.
\newblock Parsed use case descriptions as a basis for object-oriented class model generation.
\newblock {\em Journal of Systems and Software}, 84(7):1209--1223, 2011.

\bibitem{Elbendak11_thesis}
Mosa~E Elbendak.
\newblock {\em Requirements-Driven Automatic Generation of Class Models}.
\newblock PhD thesis, Northumbria University, 2011.

\bibitem{Ilieva06}
MG~Ilieva and Olag Ormandjieva.
\newblock Models derived from automatically analyzed textual user requirements.
\newblock In {\em Software Engineering Research, Management and Applications, 2006. Fourth International Conference on}, pages 13--21. IEEE, 2006.

\bibitem{Friedrich11}
F.~Friedrich, J.~Mendling, and F.~Puhlmann.
\newblock Process {M}odel {G}eneration from {N}atural {L}anguage {T}ext.
\newblock In {\em Advanced Information Systems Engineering}, pages 482--496. Springer, 2011.

\bibitem{Fliedl07}
G{\"u}nther Fliedl, Christian Kop, Heinrich~C Mayr, Alexander Salbrechter, J{\"u}rgen V{\"o}hringer, Georg Weber, and Christian Winkler.
\newblock Deriving static and dynamic concepts from software requirements using sophisticated tagging.
\newblock {\em Data \& Knowledge Engineering}, 61(3):433--448, 2007.

\bibitem{Fliedl05}
G{\"u}nther Fliedl, Christian Kop, and Heinrich~C Mayr.
\newblock From textual scenarios to a conceptual schema.
\newblock {\em Data \& Knowledge Engineering}, 55(1):20--37, 2005.

\bibitem{Capuchino00}
A~Moreno Capuchino, Natalia Juristo, and Reind~P Van~de Riet.
\newblock Formal justification in object-oriented modelling: A linguistic approach.
\newblock {\em Data \& Knowledge Engineering}, 33(1):25--47, 2000.

\bibitem{Juristo99}
Natalia Juristo, Jos{\'e}~L Morant, and Ana M~Moreno.
\newblock A formal approach for generating oo specifications from natural language.
\newblock {\em Journal of Systems and Software}, 48(2):139--153, 1999.

\bibitem{Shinde12}
Subhash~K. Shinde, Varunakshi Bhojane, and Pranita Mahajan.
\newblock Nlp based object oriented analysis and design from requirement specification.
\newblock {\em International Journal of Computer Applications}, 47(21):30--34, June 2012.

\bibitem{Ibrahim10}
Mohd Ibrahim and Rodina Ahmad.
\newblock Class diagram extraction from textual requirements using natural language processing (nlp) techniques.
\newblock In {\em Computer Research and Development, 2010 Second International Conference on}, pages 200--204. IEEE, 2010.

\bibitem{Li04}
Ke~Li, RG~Dewar, and RJ~Pooley.
\newblock Requirements capture in natural language problem statements.
\newblock {\em Heriot-Watt Technical Report HW-MACS-TR-0023}, 2004.

\bibitem{Liu04}
Dong Liu, Kalaivani Subramaniam, Armin Eberlein, and Behrouz~H Far.
\newblock Natural language requirements analysis and class model generation using ucda.
\newblock In {\em Innovations in Applied Artificial Intelligence}, pages 295--304. Springer, 2004.

\bibitem{Deeptimahanti11}
Deva~Kumar Deeptimahanti and Ratna Sanyal.
\newblock Semi-automatic generation of uml models from natural language requirements.
\newblock In {\em Proceedings of the 4th India Software Engineering Conference}, pages 165--174. ACM, 2011.

\bibitem{Deeptimahanti09}
Deva~Kumar Deeptimahanti and Muhammad Ali~Babar.
\newblock An automated tool for generating uml models from natural language requirements.
\newblock In {\em Automated Software Engineering, 2009. ASE'09. 24th IEEE/ACM International Conference on}, pages 680--682. IEEE, 2009.

\bibitem{Wahono02}
Romi~S Wahono and Behrouz~H Far.
\newblock A framework for object identification and refinement process in object-oriented analysis and design.
\newblock In {\em Cognitive Informatics, 2002. Proceedings. First IEEE International Conference on}, pages 351--360. IEEE, 2002.

\bibitem{Diaz05}
Isabel Diaz, Oscar Pastor, and Alfredo Matteo.
\newblock Modeling interactions using role-driven patterns.
\newblock In {\em Requirements Engineering, 2005. Proceedings. 13th IEEE International Conference on}, pages 209--218. IEEE, 2005.

\bibitem{Some06}
St{\'e}phane~S Som{\'e}.
\newblock Supporting use case based requirements engineering.
\newblock {\em Information and Software Technology}, 48(1):43--58, 2006.

\bibitem{Samarasinghe05}
Nayanamana Samarasinghe and St{\'e}phane~S Som{\'e}.
\newblock Generating a domain model from a use case model.
\newblock In {\em IASSE}, page 278, 2005.

\bibitem{klein03}
D.~Klein and C.D. Manning.
\newblock Accurate unlexicalized parsing.
\newblock In {\em Proceedings of the 41st Annual Meeting on Association for Computational Linguistics-Volume 1}, pages 423--430. Association for Computational Linguistics, 2003.

\bibitem{Miller90}
G.A. Miller, R.~Beckwith, C.~Fellbaum, D.~Gross, and K.J. Miller.
\newblock Introduction to wordnet: An on-line lexical database.
\newblock {\em International Journal of Lexicography}, 3(4):235--244, 1990.

\bibitem{Fellbaum10}
Christiane Fellbaum.
\newblock Wordnet.
\newblock In Roberto Poli, Michael Healy, and Achilles Kameas, editors, {\em Theory and Applications of Ontology: Computer Applications}, pages 231--243. Springer Netherlands, 2010.

\end{thebibliography}
\end{document}